# Quantum Transport in Topological Insulator Hybrid Structures——A combination of topological insulator and superconductor

OU YongXi[1], SINGH Meenakshi[2], WANG Jian[1]*

[1]*International Center for Quantum Materials, School of Physics, Peking University, Beijing, 100871, China*
[2]*The Center for Nanoscale Science and Department of Physics, The Pennsylvania State University, University Park, Pennsylvania 16802-6300, USA*

*Corresponding author (WANG Jian, email: jianwangphysics@pku.edu.cn)

Abstract:

In this paper, a brief review of the history of topological insulators is given. After that, electronic transport experiments in topological insulator-superconductor hybrid structures, including experimental methods, physical properties and seemingly contradictory observations are discussed. Additionally, some new topological insulator hybrid structures are proposed.

Contents

# 1 The discovery and properties of topological insulators

Conceptually, topological insulators (TIs) are understood to be analogous to Quantum Hall Effect (QHE) without an applied magnetic field (1, 2). In the traditional QHE, at low temperatures, in strong external magnetic fields, the electrons in a two-dimensional (2D) electron gas system transport charge along the edge of the sample. Moreover, each of these 'edge states' is associated with a definite electron momentum. Like QHE, in the Quantum Spin Hall Effect (QSHE), electrons with different spins transport charge in different directions. But unlike QHE, QSHE does not need an external magnetic field but arises from the special band structure of the material itself. Materials possessing this special band structure are known as TIs.

In TIs, the special band structure depends on two key characteristics: the spin-orbit coupling (SOC) and the time-reversal invariant symmetry (TRIS) (1, 2). In the bulk, the band structure of TIs is the same as that of a bulk insulator, i.e. with an energy gap between the conduction band and valence band. However, the bands of a TI exhibit band inversion and crossing and form Dirac cone structure, which induces a gapless metallic surface state. In the surface state, specific electronic spins are associated with specific momenta. As a result, the surface electrons travelling along certain direction have the same spin (spin-momentum locked). One consequence of this spin-momentum locking is that the surface state is protected against back scattering from non-ferromagnetic impurities and disorder. Effectively, the surface state is protected by TRIS. Unlike QHE, the surface state of TIs shows up at zero external magnetic field and room temperature.

Inspired by QHE, a 2D QSHE was predicted (3). Following this prediction, a theory discussing the necessary conditions to realize two-dimensional TIs was formulated (4). Subsequently, a specific material, HgTe quantum well, was predicted to be a true 2D TI (5) and confirmed to be such by transport experiments (6). This novel discovery led to a spate of theoretical studies in the field of TIs and predictions concerning the potential of several materials to be topological insulators. For instance, $Bi_{1-x}Sb_x$ alloy was predicted to be a three-dimensional (3D) TI (7) and was soon confirmed to be so by experiments using Angle Resolved Photoelectron Spectroscopy (ARPES) technique(8). The complex band structure of $Bi_{1-x}Sb_x$ alloy was not conducive to further in-depth experimental studies. Fortunately, other materials like $Bi_2Te_3$, $Bi_2Se_3$ and $Sb_2Te_3$ were also predicted and confirmed to be strong 3D TIs (9-11). These 'new generation' TIs have simple surface structures, which include just a single Dirac cone. This has made $Bi_2Te_3$, $Bi_2Se_3$ and $Sb_2Te_3$ the archetype TIs and these materials have attracted much attention in theoretical and experimental studies (Fig 1)[1].

Theoretical studies and experiments show many fascinating properties of TIs. For example, one of the important characteristics of TIs, the spin-momentum locked surface state, was confirmed in experiments soon after its prediction (Fig 2). Using spin ARPES, researchers measured the spin directions of the electrons on the surface and found that they were parallel to the surface and perpendicular to the momentum directions, which was consistent with theoretical predictions (12, 13). The in-plane magnetoresistance (MR) of TI thin films at different angles between the current and magnetic field directions were also studied. The observed anisotropic MR behavior found in these studies could be a result of the intrinsic spin-momentum locked property of the surface state (14). However, more experiments and analysis are necessary to confirm the spin-momentum locked transport property of surface electrons.

---

[1] All of the figures in this review are authorized by the authors.

Ideal 3D TIs have many similarities with graphene. For example, both are 2D conducting materials and have Dirac cones in their band structures (15). However, they also have several striking differences. Graphene has even Dirac cones, but TI has odd cones and shows current-induced spin polarizations in surface state. Moreover, the Fermi level in graphene naturally lies in the Dirac cone. Conversely the position of Fermi level is determined by bulk physics, not surface physics in TIs (16). Another difference between the two is that graphene loses its 2D properties as layers are added to make graphene thicker. In contrast, Molecular Beam Epitaxy (MBE) grown TI films show that TI thin films lose the 2D surface state as they are made thinner (less than 5 quintuple layers). The explanation for this is that the coupling of the two surface states opens the surface gap (17).

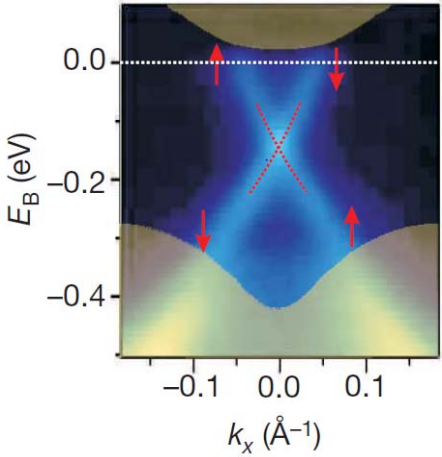
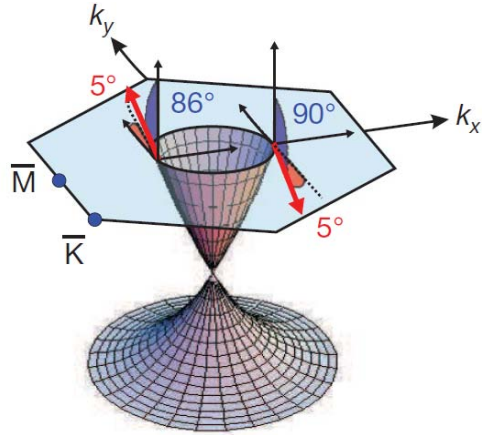

Fig 1                                   Fig 2

Fig 1. Dirac cone observed in $Bi_2Se_3$ using ARPES (Ref 13); Fig 2. Schematic of the spin-momentum locking in TI (Ref 13).

The metallic surface state of TIs is clearly seen in APRES measurements. To see this state in electronic transport measurements is more challenging. The main obstacle in the interpretation of transport results stems from the fact that the Fermi level in TIs needs not be in the bulk band gap. When the Fermi level is not in the bulk band gap, the bulk conduction states contribute to the electronic transport and the signal from the surface states may be masked. However, many experimental studies have devised ingenious ways to detect signatures of the surface state in transport measurements. Recently, some important quantum phenomena have been observed in TIs. For example, Shubnikov-de Haas (S-dH) oscillations, a common phenomenon in 2D electron gas system, were observed in TIs. Moreover, these oscillations were found to depend only on the field component normal to the surface indicating that they originate in the 2D surface state (18). Aharonov-Bohm effect produced by coherent conduction electrons was seen in $Bi_2Se_3$ nanoribbons, which is another possible evidence in transport measurements for the existence of TI surface state (19). At low temperatures an MR dip around zero field induced by 2D weak antilocalization (20-23) and resistance upturn affected by 2D electron-electron interaction (23, 24) were discovered and analyzed in TI materials. Additional efforts involving doping and top and bottom gating to separate the surface transport from bulk transport in TIs are underway.

The unique band structure and topologically protected states of TIs makes them a treasure house for exploring new physics and novel devices. The potential for exciting new physics is compounded when these materials are interfaced with other materials having quantum states like superconductors (SCs) or ferromagnets (FMs). One such combination – that of TI and SC has been the focus of much research in the last few years. One impetus for this research is an attempt to look for the Majorana fermion (MF), which was theoretically predicted several decades ago and has not yet been definitively confirmed by experiments. MF is its own antiparticle and has many properties different from other fermions (25). Some experimental setups to observe MFs have been suggested by theorists. One of these is to combine TI and SC to form a hybrid structure (see the following content). The detection of MF is not only important for the fundamental physics, but also has great potential in practical applications for example in the realization of a topological quantum computer (26, 27).

## 2 The combination of topological insulator and superconductor

1) Theoretical discussions

As mentioned above, experiments on hybrid TI/SC began as attempts to detect MFs which had been predicted to exist in the interface between TI and SC (28). Taking into account the proximity effect, a non-chiral Majorana bound state was predicted to exist in the vortices in the vicinity of the TI-SC interface.

Predictions of other novel phenomena that might occur in TI-SC hybrid structures have also been made and different experiments to realize these were proposed. For example, Liang and Kane designed a SC-2D TI-SC junction in an rf SQUID geometry (29), which could induce fractional Josephson effect (JE), i. e. $4\pi$ period in the current-phase relation, different from the $2\pi$ period in the traditional Josephson junctions (JJs). The TI-SC-TI structure was also theoretically studied with superconducting electrodes and externally applied magnetic fields (30). The results of the study suggested that, in addition to the $4\pi$ period, the MFs in the TI edges could induce an additional fractional Josephson effect. The proposed mechanism of the fractional Josephson effect was that the Cooper pairs in the middle SC split and inject into two TI sides via the Majorana states. This unconventional Josephson signature can be detected through Shapiro-step measurements.

More proposals have been made to detect MFs in different JJ structures. The transport properties of a normal metal (N)-ferromagnet insulator (FI)-superconductor (SC) junction and an SC-FI-SC junction formed on the surface of a 3D TI were theoretically studied (31). Both SC and FI may open the gap of the TI surface state. In this kind of structure, it is found that the generated Majorana mode is very sensitive to the direction of the magnetization in the FI region. This implies that the current-phase relation of the Josephson current can be tuned continuously by the component of magnetization perpendicular to the interface. Measuring the Josephson current relationship as a function of field in this geometry is then a practical way to see a signature of MFs in an electronic transport measurement. Circular JJ was also theoretically suggested as a way to observe the Majorana zero mode (MZM) (32). Fluxons occurring in this junction were predicted to be non-Abelian solitons.

The electron pairing mechanism in the proximity-induced superconducting TI was theoretically studied. By developing an effective low-energy model for the proximity effect, it was shown that the surface state of the TI would penetrate into the SC part and result in renormalization of the effective surface TI Hamiltonian (33). In this situation, p-wave pairing was predicted to occur away from the Dirac cone vertex. Close to the Dirac point, the induced *s*-wave superconductivity can be observed. Another theoretical study shows that in a (FI-d wave superconductor)/TI heterostructure, when spin-triplet pairing happens in the interface, there is no superconducting gap and Andreev reflection may not appear (34). As for spin-singlet pairing, the zero-energy surface state in the $d_{xy}$-wave SC becomes a Majorana fermion, which can be detected experimentally.

Some theoretical discussions have been focused on the conditions to detect MZM in real TI-SC heterostructures. For example, the bulk of a 3D TI penetrated by a magnetic flux can induce a gap in the surface vortex modes and destroy the Majorana mode (35). Another study analyzed the possibility of MZM in the doping-induced superconducting TIs (e.g. p-doped $Bi_2Te_3$) and pointed out that the MZM in surface state could exist only below a critical doping (36). Very recently, a prediction was made and an experimental design was suggested for studying the stability of Majorana fermions in proximity-coupled TI nanowires (37).

2) Experimental results

In comparison to the rich and mature theoretical studies discussed above, the experimental studies on TIs and TI-SC, TI-FM hybrid structures are still in their infancy. However, some recent experimental results have revealed many interesting phenomena and provided the groundwork for future research.

# I The superconductivity in topological insulators

From an experimental standpoint, the simplest technique to study how TIs interact with SCs is to detect superconducting signals in TIs. Several transport experiments at low temperatures (38-41) and scanning tunneling spectroscopy (STS) studies in ultrahigh vacuum (40) have been done to explore proximity induced superconductivity in TIs.

A proximity-induced superconducting signal was observed in $Bi_2Se_3$ thin layers (38), which were exfoliated from $Bi_2Se_3$ single crystal. Al/Ti electrodes were patterned on $Bi_2Se_3$ using electron-beam lithography (EBL) technique. Below $T_C$, a supercurrent (Ic ∼ 200nA) was observed between two superconducting electrodes separated by ∼ 400nm. By tuning the gate voltage, researchers successfully changed the value of the critical current, which showed an ambipolar gate-voltage-dependent behavior. This observation is consistent with the reported Dirac cone dispersion relationship of the surface state.

Concurrently, a separate study reported the superconducting proximity effect in $Bi_2Se_3$ nanoribbons (39). The nanoribbons were grown by gold catalyzed vapor-liquid-solid (VLS) mechanism and superconducting tungsten electrodes were deposited on the ribbons using focused ion beam technique. Below $T_C$, a supercurrent was observed between two superconducting electrodes with an effective contact separation of 580 nm (Fig 3).

A signature of superconductivity below 2–3 K has also been seen in $Bi_2Se_3$ and $Bi-Bi_2Te_2Se$ thin films (40) via point contact conductance measurements. The observed critical current dips at about ±2mV and a zero-bias conductance peak were attributed to proximity-induced local superconductivity in the films due to small amounts of superconducting Bi inclusions or segregation on the surface of TI films.

The observation of JE in TI-SC junctions has also been reported (41). Polycrystalline $Bi_2Te_3$ flakes and superconducting Nb electrodes were used to form JJs. At 1.6 K, a supercurrent with a critical value of 1.8 μA was detected with 50 nm electrodes separation. Fraunhofer pattern and Shapiro steps, two characteristics of JE, were observed (Fig 4).

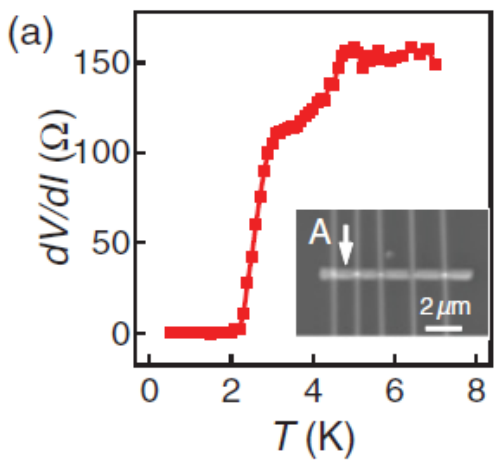
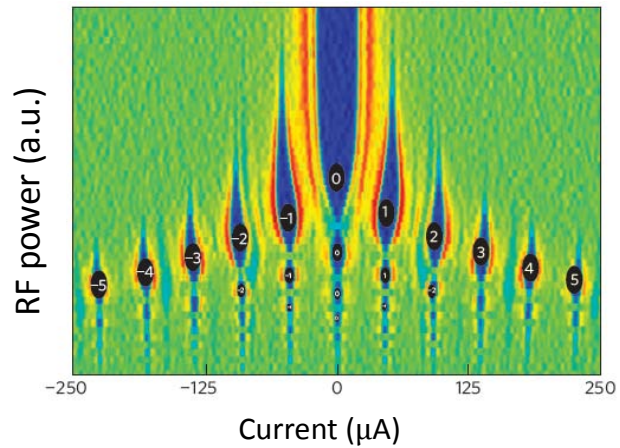

Fig 3            Fig 4

Fig 3. Proximity induced superconductivity in a $Bi_2Se_3$ nanoribbon (Ref 39); Fig 4. Shapiro steps in a $Bi_2Te_3$ Josephson junction (Ref 41).

Other than transport measurements, STS is an effective way to detect superconductivity in TIs. Recently, successful fabrication of TI-SC double-film hybrid structures grown by MBE technique has been reported (42). The local density of states (LDOS) was measured via STS to detect the superconducting signal in the TI film. Similar to other STS measurements, the observed differential conductance curves show a gap around zero bias and two peaks at certain bias voltages, supporting the idea of proximity-induced superconductivity in TIs.

## II Bulk channel or surface channel?

After the confirmation of the superconducting proximity effect in TIs, a natural question to ask is whether the bulk or surface modes are supporting the supercurrent. Ideal TIs have a gapped insulating bulk and the metallic surface states. However, in real samples, the Fermi surface may not be in the gap and the bulk contribution to electronic transport cannot be ignored. The carriers in bulk come from impurity band induced by disorders or vacancies in the crystal (38, 43). Initial observations of S-dH oscillations in TIs were open to discussions on bulk or surface states contribution. However, now it is widely accepted that S-dH oscillations rooted in the 2D surface state really exist in TI materials (18, 41, 44). Taking ideas from the S-dH results, we can think about answering the question of whether the superconducting proximity effect is supported by the bulk or the surface states.

The key to answer this question is to exploit the difference between the surface states and the bulk state. For example, gate voltage can be used to tune the Fermi level to see the dependence of the supercurrent on this. Ambipolar dependence of the supercurrent on the gate voltage has been observed (38). This was attributed to the surface state inside the gap of the bulk bands and proved that the surface channels (Dirac fermions) participate in the supercurrent transport.

Generally, in order to detect the superconducting proximity effect, the sample size should be smaller than two characteristic lengths, the thermal length and the phase-breaking length. TI nanoribbons with different superconducting electrode separations were measured (39). In these SC-TI-SC junctions, the SC separations were larger than the two characteristic lengths of the bulk but shorter than those of the ballistic surface. The observed supercurrent supports the idea that the proximity-induced superconductivity in TI ribbons is likely to have its origins in the surface state.

Furthermore, Veldhorst *et al.* measured the dependence of critical supercurrent on temperature and junction size (41). The results of their experiment were consistent with the clean limit description (mean free path ＞ coherence length) instead of the dirty limit (mean free path ＜ coherence length). The agreement with the clean limit suggests ballistic transport, which in turn implies that the observed Josephson supercurrent mainly spreads through the ballistic surface channel instead of the diffusive bulk channel.

Although these experimental results clearly show that the TI surface contributes to the supercurrent transport in TI-SC junctions, they do not rule out bulk contribution. Yang *et al.* fabricated Pb-$Bi_2Te_3$-Pb lateral and sandwiched junctions to study the superconducting proximity effect (47). The observations suggest that the bulk of $Bi_2Te_3$ flake in the junction becomes superconducting too. Moreover, STS experiments by Wang *et al.* also indicate the TI bulk can be superconducting due to proximity effect (42). To summarize, both surface and bulk contributions are critical in explaining the experimental results in SC-TI-SC systems.

## III Two unexpected phenomena

It is worth discussing two unexpected phenomena observed in transport experiments on SC-TI-SC structures. One is the resistance enhancement below the critical temperature (Tc) (45) and the other is the unusual Fraunhofer pattern of these structures (46). These two phenomena may help us in understanding the special properties of the superconducting surface state of TIs.

Proximity induced superconductivity in TIs has already been observed and reported (38, 39, 48, 49). However, a study on TI thin films contacted by SC electrodes found an unusual resistance increase when TI samples were cooled down to the critical temperature of the superconducting electrodes (45). The $Bi_2Se_3$ thin films samples used in the study were grown by MBE method on sapphire and high-resistance silicon substrates. Hall results suggested that the films were of high quality. Control experiments were done by using different superconducting electrodes (In, Al and W), different separation lengths of electrodes (1mm and 1μm) and different thicknesses (5QL and 200QL) of TI thin films. All results showed that the resistance increased just below the Tc of the electrodes. The increased resistance below Tc can be more than ten times larger than the normal resistance just above Tc (Fig 5). An applied external magnetic field suppresses the increase in resistance. These features could not be explained by the conventional Blonder-Tinkham-Klapwijk (BTK) theory for proximity-induced superconductivity after consideration of the contact resistance. Therefore, it was suggested that the special helical structure of the TI surface might be contributing to this counter-intuitive behavior. In the SC electrodes, the electrons in a Cooper pair have spins aligned antiparallel to each other whereas on the TI surface the spin-momentum locked state requires electrons to have spins aligned parallel to each other for charge transport. This incompatible spin pairing in SC and TI surface might induce the resistance increase below Tc. To definitively establish this idea, more theoretical and experimental studies are needed.

The other unexpected phenomenon was seen in the measurement of magnetic diffraction pattern. Some experiments have shown that Josephson current could exist in TI junction and induce a Fraunhofer pattern akin to that seen in regular JJs. Initial measurements showed the usual Fraunhofer pattern (41, 47) including trivial deviations that came from the size effect (45). However, in one of the experiments, scientists reported an anomalous Fraunhofer pattern (46). The magnetic field related to the first minima of Ic was five times smaller than the predicted value and the minima of Ic were not equally spaced as a function of the magnetic field (Fig 6). This was significantly different from the traditional Fraunhofer pattern. Performing control experiments using graphite instead of $Bi_2Se_3$, scientists got rid of possible geometry reasons and attributed the result to the TI–SC contact. In order to explain this phenomenon, it is necessary to consider both bulk current-phase relationship of the JE and an unconventional current-phase relationship from the surface states. Simulations of the magnetic diffraction patterns taking into account both the bulk and the surface state were successful in reproducing the experimental results.

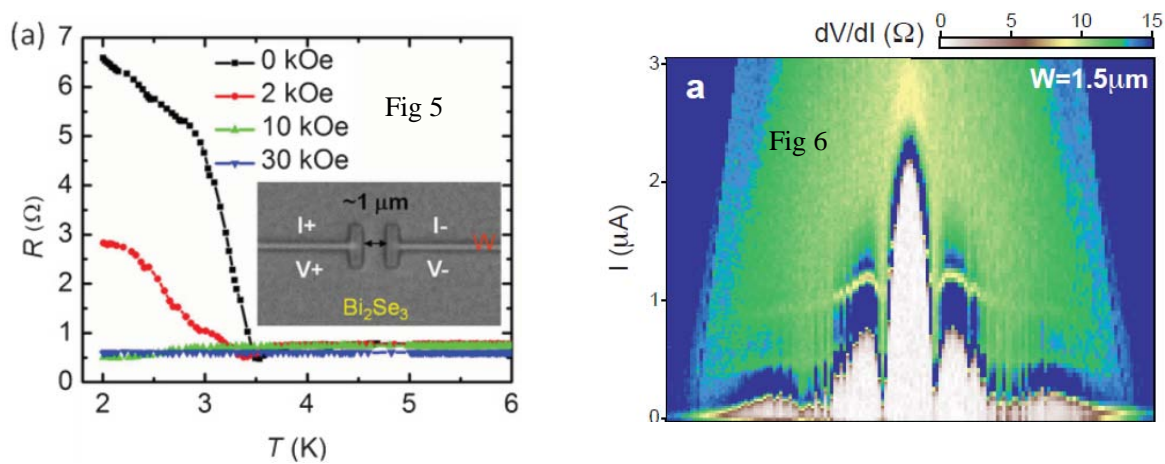

Fig 5. Unexpected resistance increase in $Bi_2Se_3$ thin films (Ref 45); Fig 6. Unusual magnetic diffraction pattern in a $Bi_2Se_3$ Josephson junction (Ref 46).

These two unexpected phenomena suggest that there may be many undiscovered novel properties of the TI-SC hybrid structures. The differences between theoretical predictions and experimental results are stimulating for further studies to gain a better understanding of this fascinating system. One example of an exciting possible future experiment is that the theoretically predicted current-phase relationship (4π period) (29) remains to be observed experimentally (46, 49).

## IV The superconducting signal under bias voltage

In a SC-TI hybrid system, a peak in the vicinity of zero bias in the differential conductance curves is observed at temperatures below Tc. As the temperature decreases, the height of this zero bias conductance peak (ZBCP) increases and its width decreases (40, 48). Applying an external magnetic field suppresses the ZBCP. Control experiment using graphite (no observation of ZBCP) suggests that the ZBCP is relevant to the surface state of TI. It is believed that the ZBCP was related to unconventional superconductivity in the TI-SC interface. It is a possible signature of Majorana fermions.

Other than the measurements under zero bias, superconducting signals in TI-SC structures under non-zero bias also provide much information about the physics of the system. These signals are usually sensitive to temperature and external field.

Andreev reflections (ARs) can be observed in the interface between TI and SC in transport measurement (50). When an electron from the normal side is injected into the SC side, a hole simultaneously reflects from the interface, causing the differential conductance to be twice its usual value. Multiple Andreev reflections occur in the SC-N-SC structure and the positions of the differential conductance peaks depend on the bias value with the peaks being located at $V_{bias}=2\Delta/ne$ where n is an integer, e is the electronic charge and $\Delta$ is the superconducting gap. ARs and multiple ARs have been observed in TIs (38, 39, 47), though the experimental results are not completely consistent with theoretical predictions (Fig 7).

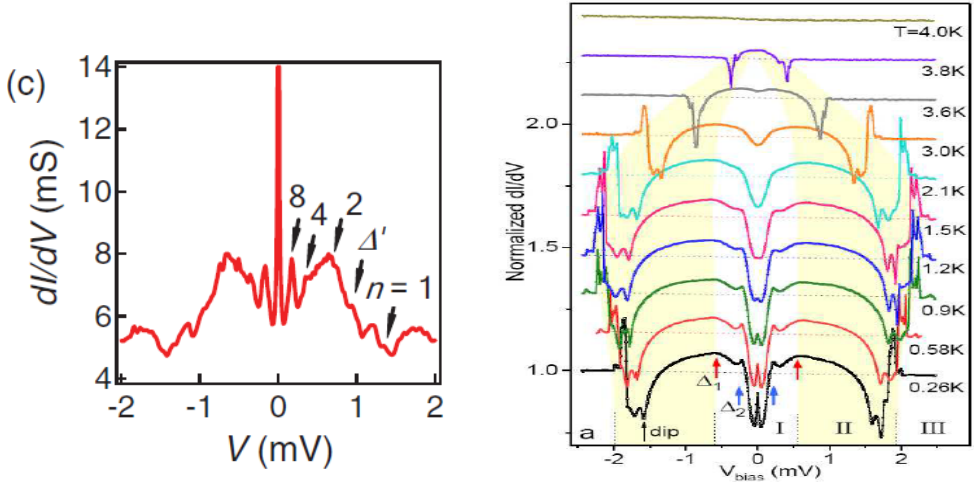

Fig 7                               Fig 8

Fig 7. Multiple Andreev reflections in SC-TI-SC nanoribbon system (Ref 39);

Fig 8. Zero bias conductance peak and double gap structure of a Sn-$Bi_2Se_3$ junction (Ref 48).

Measurement of differential conductance under bias in SC-TI hybrid structures is critical because it reveals important information not seen in other transport measurements, for example, the superconducting energy gap. Two energy gap signals in TIs were reported (48). The double gap structure was observed in the conductance curves under voltage bias. One of the gaps had its origin in the SC electrodes (Fig 8). The second gap was seen in most TI samples but disappeared in graphite sample. Other researchers reported symmetric dips in differential resistance versus current bias curves (46), which might be related to the double gap behavior mentioned above. Yang *et al.* suggested that the double gap structure is induced by a new superconducting phase occurring in the TI-SC interface.

## V The mystery of $I_CR_N$

As mentioned earlier, JJs of TIs have been fabricated and Josephson supercurrent has been observed. The properties of these JJs are however, somewhat different from those of traditional JJs. In JE, $I_CR_N$ (where Ic is the critical current under zero bias and zero magnetic field and $R_N$ is the junction resistance in the normal state) is a useful value to estimate how different a JJ is from an ideal JJ. Ideal JJs have $I_CR_N$ independent of the size of junctions, but dependent on the temperature and the superconducting energy gap of electrodes. The value of $I_CR_N$ can be estimated by $\Delta/e$ (50). It is interesting that in almost all reported experiments on JJs of TIs, the $I_CR_N$ value is much smaller than this prediction (38, 39, 41, 48, 49). Yet in the control experiment using graphite, the value is more consistent with theoretical prediction. These results seem to suggest that the small $I_CR_N$ is related to the special properties of TIs.

One possible explanation is that the much smaller $I_CR_N$ is due to the difference between the surface and the bulk of the TI (41). Experiments have suggested that proximity-induced supercurrent exists mainly in the surface channel (38, 39, 41). If the surface state is the dominant contribution to supercurrent transport, $I_C$ should be estimated using the surface channel. However, when the system is in normal state (non-superconducting state), the transport contribution from the bulk is important (38). As a result, when we measure the normal resistance, we actually measure the surface resistance shunted by bulk resistance. It is a small value because the bulk shunt strongly decreases $R_N$. Therefore, the observed $I_CR_N$ value in SC-TI-SC system is smaller than the theoretical prediction.

Another explanation attributes the small $I_CR_N$ to the formation of special energy structure in the SC-TI JJs (46). In this model, a 1D Majorana line along the JJs is constructed. This linear structure shows quantized energy levels. The lowest level around zero point has energy $E_C = hv_{ex}/2W$, in which $v_{ex}$ is the velocity of carriers in the line and W is the width of the junctions. It was assumed that the supercurrent could spread only in the zero level. As a result, $I_CR_N \propto E_C/e \propto 1/W$, instead of $\propto \Delta/e$ (Fig 9), and is directly related to the value of W. Junctions with different widths were measured and it was confirmed that the $I_CR_N$ value becomes smaller with larger width.

Identifying and proving the real reason behind this effect (small $I_CR_N$) remains an open challenge. Without doubt, the resolution of this mystery will help us in understanding the interaction between the SC phase and the topological quantum state.

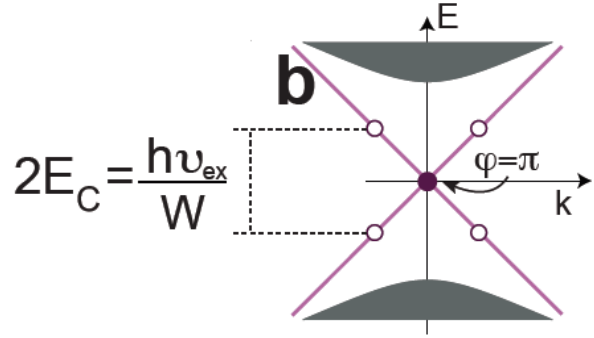
Fig 9

Figure: 9. Majorana line model in TI JJs (Ref 46).

## VI The influence of geometry

In transport measurements, the sample geometry and size could influence the results, especially in nanoscale systems. It is therefore useful to discuss some geometrical considerations that may affect the outcome while performing transport measurements on these systems.

Point contact is one way to obtain transport properties of TIs. Gold tips have been served as point contact junctions in measuring TI thin films (40). As shown in experiments, the configuration of the contact area could be complex. Using this contact technique, differential conductance dips or peaks at different biases were observed, which suggested occurrence of local superconductivity.

In most experiments, superconducting electrodes were directly deposited (via evaporation, sputtering or focused ion beam technique) on the TI nanostructures (thin films, flakes etc.) to form TI-SC hybrid systems. TI-SC junctions with different dimensions were used to measure magnetic diffraction patterns (47). It was found that the measurement results were more consistent with ideal Fraunhofer pattern when the SC electrodes spanned the entire width of the TI sample. If the electrodes only span a part of the TI flake, e.g. if they are located in the middle area away from the edges, the observation would deviate from ideal pattern, showing lower critical current peaks (Fig 10). This deviation is attributed to non-uniform spatial distribution of the supercurrent.

The size of junctions may also affect other behavior. Hysteretic behavior of I-V curves in the superconducting transition range was reported in Pb-$Bi_2Te_3$-Pb JJs (47) (Fig 11), which has not been observed in other experiments (46, 49). Generally, JJs can be divided into two types according to the resistively and capacitively shunted junction (RCSJ) model: overdamped and underdamped (50). For underdamped JJs, hysteretic I-V curves can be observed. This usually happens in the case where the junction resistance or capacitance is large. However, in Pb-$Bi_2Te_3$-Pb system (47), the bridge-type JJs have small capacitance. Also, the normal state resistance is small. The observation of hysteretic I-V curves in this system is therefore unexpected. After comparing different experiments, it was found that the probability of observing hysteretic I-V curves increases with the separation of the superconducting electrodes. Taking the unusual increase of resistance (45) below Tc into consideration, further control experiments in JJs are necessary to clarify the mechanism of hysteretic IV behavior. Performing measurements with different superconducting contact separations, different superconducting electrodes, and different TI materials may be the key to understand this hysteretic phenomenon.

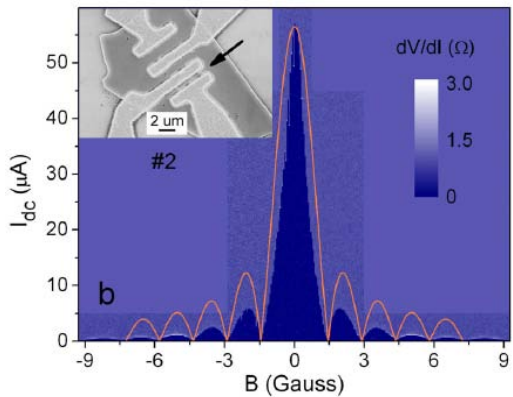 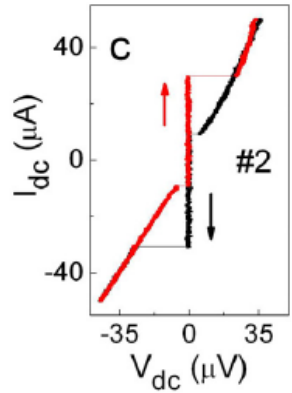

Fig 10                                   Fig 11

Fig 10. Deviated magnetic diffraction pattern (Ref 47); Fig 11. Hysteretic behavior of I-V curve of the Pb- $Bi_2Te_3$-Pb JJs (Ref 47).

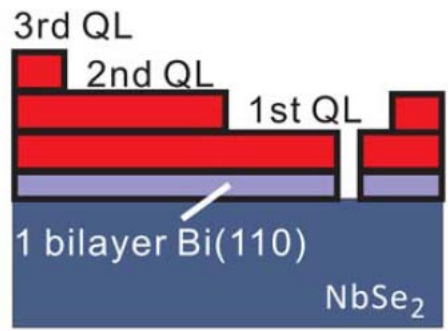

Fig 12

Figure: 12. Schematic of TI-SC double-film structure (Ref 42).

Recently, Wang *et al.* reported a new kind of measurement geometry, epitaxial TI-SC double-film structure (42) (Fig 12). In order to overcome the lattice mismatch between the superconducting $NbSe_2$ substrate and $Bi_2Se_3$ thin films, a Bi bilayer film was first deposited on the $NbSe_2$ substrate and then the $Bi_2Se_3$ thin films were grown by MBE. The *in situ* STS and ARPES measurements confirmed the coexistence of superconductivity and TI surface state in this structure. This geometry offers a new platform to study the interplay between TI surface state and superconducting state. For example, the direct detection of the proximity-induced superconducting vortices on the TI surface (35), the Majorana zero mode, and possible magnetic signal induced by spin magnetic torque in the transport measurement (51).

VII Magnetic effect

The possibility of magnetic effects in TIs has been the focus of research from a fundamental physics perspective as well as for potential applications. For example, the long sought magnetic monopole is predicted to exist in TI surface state (52). With proper doping, TI can induce magnetic order and quantized anomalous Hall effect (53). The spin-orbit coupling also makes TI a candidate for spintronics (16), which exploits the tunable spin properties and has potential in electronics industry. Although the promising field of magnetic effects in TIs is still in its infancy, some results have already emerged.

Magnetoresistance (MR), i.e. resistance as a function of externally applied magnetic field, is an important signal to study quantum transport in TI-SC structures. In these systems, the MR exhibits periodic oscillations with 1/H (where H is magnetic field), i.e. S-dH oscillations (18, 38, 41). However, some researchers observed other MR oscillations in TIs. Zhang *et al.* used tungsten electrodes to measure the MR of $Bi_2Se_3$ nanoribbons and observed MR oscillations with the applied field perpendicular to the plane of the sample when W electrodes became superconducting (39). Unlike S-dH oscillations, the period of the observed oscillations was proportional to H, rather than 1/H. Moreover, the amplitude of the oscillations changed nonmonotonically with changing temperature. The oscillations disappeared in parallel field and also when the temperature was increased above the Tc of the tungsten electrodes. The disappearance in parallel field rules out the Aharonov-Bohm effect as a possible explanation for these MR oscillations. The 'Weber blockade' vortex model was suggested to explain these observations (54). It was suggested that Pearl vortices may be induced on the TI surface by superconducting proximity effect.

The periodic MR oscillations in the SC state are mainly related to the size of the samples. The following behavior, observed in TI films contacted with SC electrodes, may have a more intrinsic link with the SC-TI interface. Wang *et al.* reported unusual MR behavior in TI films contacted by superconducting electrodes (45). When the indium electrodes became superconducting, negative MR appeared from zero magnetic field to higher field. The negative MR persisted even when the applied magnetic field exceeded the critical field of superconducting electrodes, but, could not be observed with nonsuperconducting electrodes. Careful measurements revealed hysteretic behavior of MR, reminiscent of ferromagnets, around zero field.

VIII What happens at the interface?

All of the experimental results discussed above suggest the important role of TI-SC interface, which may not only contain the predicted MFs, but also be an exciting system to explore a plethora of novel physics. Theories and experiments have pointed out that, in the TI surface state, the spin and momentum are locked causing electrons transporting charge in one direction to have spins aligned parallel to each other.

The BCS theory states that the two electrons in a Cooper pair in an s-wave superconductor have different spin directions. However, spin-triplet Cooper pairs, which do not require the spins of the constituent electrons to be anti-parallel, have also been predicted and confirmed. This spin-triplet (p-wave) pairing has been predicted to exist in superconducting TIs (33). In the TI-SC interface, spin-triplet pairing electrons may come into being because of the proximity effect and current-induced spin polarization. The SC signature from the proximity-induced superconducting region in the TI is different from the SC electrode. Also, the unusual JE seen in these systems may be due to the triplet superconductivity. The possibility of spin-triplet superconductivity in the SC-TI interface may therefore be related to many unexpected experimental observations mentioned above, such as double gap structure (48), unusual signals of conductance curves (39), hysteretic behavior (47) and the unusual JE (46).

Thus, understanding the properties of the TI-SC interface is the key factor in unraveling the transport studies in TI-SC hybrid structures. The answer to the question 'What happens at the interface?' is crucial to understanding the many mysteries of TI-SC hybrid structures.

## 3   Other topological insulator hybrid structures

As described above, the combination of TI with traditional SC attracts much research interest.   Concurrently, great progress is also being made in the search for new topological materials (not only TIs) and new superconductors. Herein is a brief introduction to some of these novel materials, which may show us completely new aspects of TI hybrid structures in the future.

1)  Topological superconductors and topological semimetals

Topological superconductors (TSCs) are a new kind of superconductors which were predicted by topological band theory (2). TSCs have a superconducting gap in the bulk but no gap in the surface (55). From the view of topological structure, TSC and TI are similar. It was also predicted that MZM could exist on the surface of TSC. Some calculations have pointed out how to produce such kind of superconductors (56). Today, $Cu_xBi_2Se_3$ is thought to be a possible candidate for TSC (57).

Topological semimetal is a new state of material predicted by first principle calculations (58). Researchers found that, when $HgCr_2Se_4$ was in the ground state, band crossing would occur in its surface. Spin-up and spin-down electrons would be in different energy bands, which would induce a nontrivial semimetallic state. Some novel properties were predicted for this state. For example, it has a non-closed Fermi surface (Fermi arcs) and the anomalous quantum Hall effect may exist in this state.

2) New topological insulators and superconductors

Although in most of the current experiments, $Bi_2Se_3$ and $Bi_2Te_3$ are used as TI samples, researchers are continuing to look for new members of the TI family. Half-Heusler compounds (59), $(M_3N)Bi$ (M=Ca, Sr and Ba) (60) and $Ag_2Te$ (61) are materials predicted to be new TIs by first principle calculations. These new TIs have different properties from $Bi_2Se_3$ etc., e.g. the Dirac cone in their surface state could be anisotropic.

In ultrathin superconducting films, superconductivity is closely related to the film thickness and the influence of the substrate (62, 63). Moreover, experimental studies have shown the coexistence of the magnetic properties and superconductivity in 2D interfaces, which could help us broaden our understanding of traditional SC (64, 65). Recently, Fe-based high Tc superconducting thin films were successfully fabricated (66). Combining these new superconducting thin films with new TIs to form hybrid structures will reveal a variety of novel phenomena and aid in the understanding of the physics of these systems.

## Acknowledgements

This work was supported by the National Basic Research Program (NBRP) of China (No. 2012CB921300), National Natural Science Foundation of China (No. 11174007) and the Penn State MRSEC under NSF grant DMR-0820404. We are grateful to Li Lu, Lan Yin, M Veldhorst, JR Williams, DG Gordon, Liang Fu, MZ Hasan, DM Zhang, and Dong Qian for their permission to use their figures or helpful discussions.